\documentclass[sn-mathphys-num,Numbered]{sn-jnl}

\usepackage{soul}
\usepackage{newtxtext,newtxmath}
\usepackage{graphicx}%
\usepackage{subcaption}%
\usepackage{multirow}%
\usepackage{amsmath,amsfonts}%
\usepackage{amsthm}%
\usepackage{mathrsfs}%
\usepackage[title]{appendix}%
\usepackage{xcolor}%
\usepackage{textcomp}%
\usepackage{manyfoot}%
\usepackage{booktabs}%
\usepackage{algorithm}%
\usepackage{algorithmicx}%
\usepackage{algpseudocode}%
\usepackage{listings}%
\usepackage{setspace}
\usepackage{doi} 
 
\theoremstyle{thmstyleone}%
\theoremstyle{thmstyletwo}%

\theoremstyle{thmstylethree}%
\renewcommand{\vec}{\mathbf}
\definecolor{darkgreen}{rgb}{0.0, 0.5, 0.0}
\newcommand\persec[0]{$\cdot$s$^{-1}$~}
\newcommand\permcubed[0]{$\cdot$m$^{-3}$~}

\begin{document}

\title[Article Title]{Mixing and sharpening at the interface of a two-layer fluid forced by random jets}


\author*[1,2]{\fnm{Noé} \sur{Clavier}}\email{noe.clavier@ista.ac.at}

\author[3]{\fnm{Hugo} \sur{Pradel}}\email{hugo.pradel@ens-lyon.fr}

\author[3]{\fnm{Romain} \sur{Volk}}\email{romain.volk@ens-lyon.fr}

\author[3]{\fnm{Mickaël} \sur{Bourgoin}}\email{mickael.bourgoin@ens-lyon.fr}

\author*[3,4]{\fnm{Yvan} \sur{Dossmann}}\email{yvan.dossmann@ens-lyon.fr}

\affil*[1]{\orgdiv{Département de Physique}, \orgname{École Normale Supérieure PSL}, \orgaddress{\street{24, rue Lhomond}, \city{Paris}, \postcode{75005}, \country{France}}}

\affil[2]{\orgname{Institute of Science and Technology Austria}, \orgaddress{\street{Am Campus 1}, \city{Klosterneuburg}}, \postcode{3400}, \country{Austria}}

\affil[3]{\orgdiv{Laboratoire de Physique}, \orgname{École Normale Supérieure de Lyon}, \orgaddress{\street{46 allée d'Italie}, \city{Lyon}, \postcode{69007}, \country{France}}}

\affil[4]{\orgdiv{LEMTA}, \orgname{Université de Lorraine}, \orgaddress{\street{2 avenue de la Forêt de la Haye}, \city{Vandoeuvre-lès-Nancy}, \postcode{54500}, \country{France}}}



\abstract{Understanding mixing at density interfaces is essential for predicting transport in stratified environmental flows. Laboratory studies have mostly relied on steady, spatially uniform forcing, whereas turbulence in nature is intermittent and heterogeneous. Here, we present experiments on a two-layer salt-stratified fluid forced by random turbulent bursts generated with a randomly actuated synthetic jet array (RASJA). Density fields are recorded with the light attenuation technique, allowing us to resolve the interface evolution. We measure that the upward velocity of the interface decreases with the density jump, in agreement with the power-law found in previous oscillating-grid studies. At large density differences, the interface sharpens during mixing, contrary to the smaller density jump case. Background potential energy analysis demonstrates irreversible mixing in both cases, with comparable energy changes. These results extend classical laboratory observations to a more isotropic forcing, offering new insights into the dynamics of mixing in geophysical settings. }




\maketitle

\section{Introduction}\label{sec:intro}

Modelling the properties of turbulence and mixing under external forcing and initial stratification has been the subject of considerable research since the 1960s \citep{caulfield_layering_2021}. A major challenge lies in the faithful representation of small-scale processes and their effects on turbulent mixing, such as internal wave breaking and convective instabilities, which must be parametrised in global circulation models \citep{deLavergne2020,Mashayek2021,McDougall2021}. Among the large number of physical quantities that are evaluated to describe stratified turbulent mixing, the definition of several of them are still debated, such as the seemingly straight-forward mixing efficiency \citep{davies_wykes_meaning_2015}. The realisation of fully-controlled turbulent experiments and direct numerical simulations or the combination thereof allow to investigate the dynamics of mixing over a wide range of controlling dimensionless parameters, providing useful insights on the understanding and parametrisation of ocean-scale processes \citep{dossmann_mixing_2017,couchman_mixing_2023,petropoulos_prandtl_2024,sajeev_how_2025}.

A model stratification is the two-layer fluid, a system composed of a bottom, denser layer and a top, lighter layer separated by a narrow density interface. The stability of such a two-layer fluid submitted to a continuous mixing event was first studied in the experiments of \citet{rouse_diffusion_1955}, in which the interface was disturbed by the oscillation of a grid at the bottom of the system. The experiments were latter reproduced by \citet{hopfinger1976spatially} and \citet{gostiaux_diffusion_2013}, the latter exploiting the PLIF technique to compare the evolution of the density profile to a statistical mechanics approach prediction. The stability of stratified layers separated by narrow interfaces in these experiments have been attributed to the Phillips-Posmentier mechanism \citep{phillips_turbulence_1972,posmentier_generation_1977} which relies on a postulated non-monotonic relation between the buoyancy flux (resp. mixing efficiency) and the stratification described by the Brunt-Väisälä frequency $N = \sqrt{-g(\partial_z\rho)/\rho}$ (with $\rho$ the fluid density and $-g\hat{\vec{z}}$ the gravity). Later experiments investigated further the layer formation and stability and their relation to the turbulence source and position relative to the stratification \citep{ruddick_formation_1989,park_turbulent_1994,holford_turbulent_1999}. The dynamical structures (vortex shedding) involved in the formation of layers and their dependence on the type of grid forcing have been reviewed by \citet{thorpe_layers_2016}. Mode-1 internal waves have been identified as a source for the progressive development of density staircases in an initially linear stratification in laboratory experiments \citep{dossmann_mixing_2017}. The use of the light attenuation technique to measure density fields $\rho(\vec{r}, t)$ highlighted that vertical shear gradients control the layering process: the maximum IW-induced shear leads to the formation of a central pair of staircases, followed by secondary and tertiary staircases pairs formation above and below the first pair.

However, in all these experiments stratified layers emerge from a regular steady forcing inducing turbulence. The question of whether, and how layers emerge from intermittent, heterogeneous turbulence bursts closer to oceanic conditions is still unexplored, as well as the mixing efficiency of such events. The present work addresses these two questions by reporting light attenuation technique (LAT) and conductivity measurements in a two-layer fluid forced by a random turbulence field in the bottom phase. Contrary to most previous works where turbulence is generated by an oscillating grid or moving rods \citep{rouse_diffusion_1955, hopfinger1976spatially, ruddick_formation_1989, park_turbulent_1994, thorpe_layers_2016}, here we use a set of randomly-actuated synthetic jets (RASJA) located at the bottom of the tank. Such a forcing proved efficient to produce more homogeneous, isotropic turbulence \citep{variano_random-jet-stirred_2008}. The results reported in this paper extend previous observations to this more realistic flow, and as such are a step towards the generalisation of laboratory studies to environmental flows.

We describe the setup and the density measurement techniques in section~\ref{sec:meth}. Section~\ref{sec:displ} shows how the density interface is displaced up under mixing and relate its velocity to the controlled density gradient at the interface. Under certain conditions we observe that the sharp interfaces become even sharper under mixing and relate this observation to the theory of layering in section~\ref{sec:lay}. The potential energy budget and mixing efficiency are discussed in section~\ref{sec:energy}. Concluding remarks are exposed in section~\ref{sec:conc}.

\section{Experimental methods}\label{sec:meth}

\subsection{Generation of stratified turbulence}

\begin{figure}
    \centering
    \includegraphics[width=0.5\linewidth]{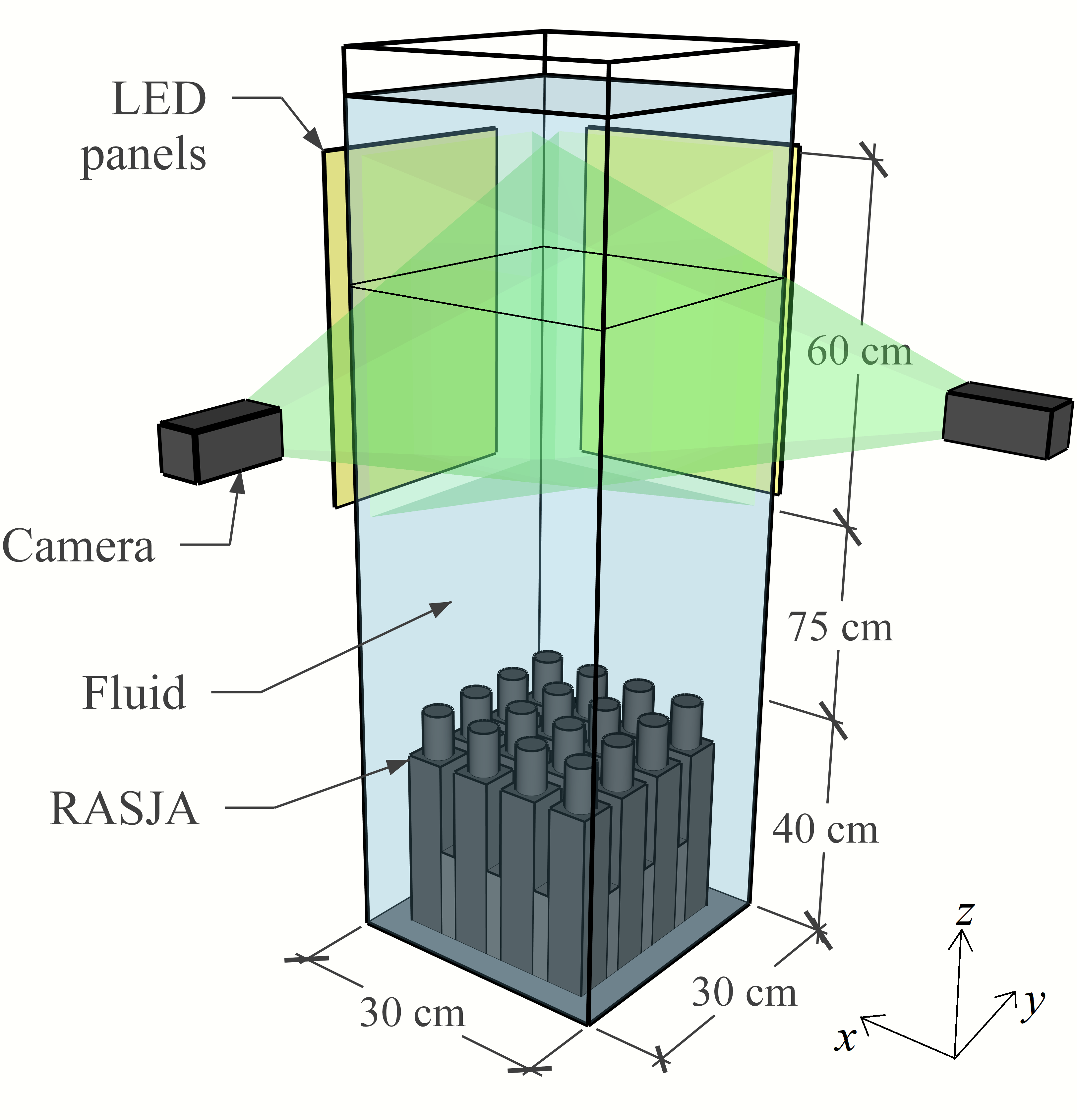}
    \includegraphics[width=0.25\linewidth]{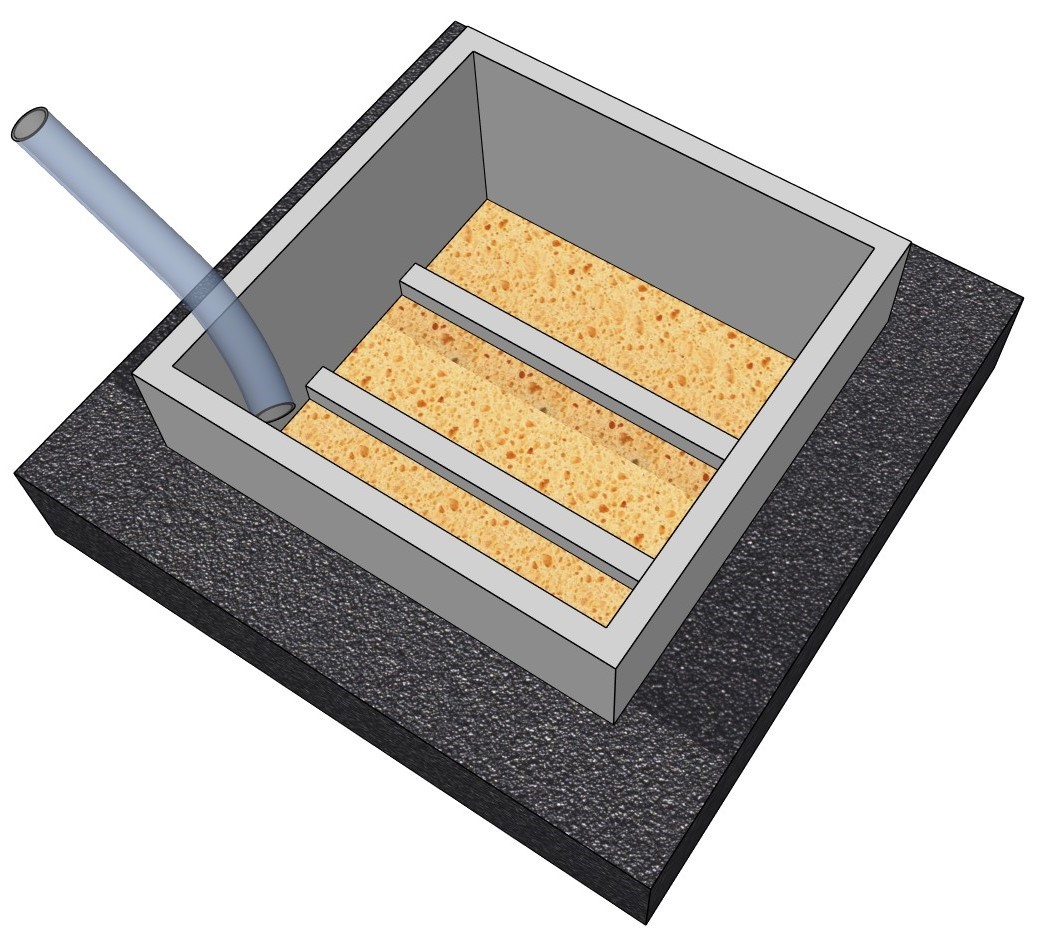}
    \caption{\textbf{Left:} Schematic of the experimental setup. \textbf{Right:} Schematic of the floating device used to generate a two-layer stratification.}
    \label{fig:setup}
\end{figure}

Experiments are carried out using the water tank and turbulence-forcing system depicted in Figure \ref{fig:setup}. The tank is 160-cm high and has a square horizontal section of inner width 30 cm. The bottom of the tank is closed, whilst the top is opened. The tank is filled up with tap water which can be pumped in and out from two holes drilled through the bottom face. Density is varied through the concentration in sodium chloride. In order to create a two-layer stratified fluid, we first fill the tank up to the desired level with tap water in which sodium chloride is dissolved. Then, the second layer is created using a specially-designed floating device to ensure a smooth filling of the upper layer without perturbing the interface. (Fig. \ref{fig:setup}). It is made of a 3D-printed structure in the shape of a square reservoir, surrounded by plastic foam to ensure buoyancy. The bottom face of the float consists in compressed sponges. The reservoir is filled with fresh water which flows very slowly through the sponge. The resulting interface is typically 1 cm thick.

At the bottom of the tank, 16 bilge pumps (Rule 25DA 500GPH by Xylem) are laid out in a $4\!\times\!4$ grid. The pumps are all independently controlled by 16 relays connected to a Raspberry-Pi controller and powered with a common 9 V voltage source. The exit pipes of the pumps are equipped with check valves. They are all parallel—pointing upwards—and the nearest-neighbour nozzles are 7.5 cm away. This choice of spacing ensures planar symmetry of the forcing with respect to the wall—the outer nozzles-to-wall distance is half the inter-nozzle distance—which tends to cancel out the stress tensor at the boundaries and suppress secondary mean flows \citep{fernando_note_1993}. The pumps are independently and randomly actuated. Suction occurs near the bottom surface of the tank and always simultaneously with the jets, which are therefore called ``synthetic jets" (because of the zero net mass flux across any closed surface surrounding a pump). Such a forcing is commonly referred to as RASJA (for Randomly Actuated Synthetic Jet Array) and it proves to produce relatively homogeneous and isotropic turbulence in $xy$ planes sufficiently far downstream from the jets and in a homogeneous fluid \citep{variano_random-jet-stirred_2008}. We set $z=0$ at the nozzles' ends (40 cm above the bottom face), $z=H$ at the tank's top and $x=y=0$ in the middle of the tank. The random actuation of the jets can be performed in various ways, but \citet{variano_random-jet-stirred_2008} found that the following scheme ensures the most homogeneous, isotropic turbulence: (i) for each pump, a random power-on duration is drawn for each pump according to a gaussian distribution of mean $\tau_\mathrm{on}$ and width $\sigma_\mathrm{on}$ (ii) after this time, a power-off duration is similarly drawn with parameters $\tau_\mathrm{off}$ and $\sigma_\mathrm{off}$ (iii) the process is repeated. \citet{laplace_etude_2022} extensively characterised the flow resulting from this forcing procedure in the present setup, in pure water. We set accordingly $\tau_\mathrm{on} = 4$ s, $\tau_\mathrm{off} = 28 $ s and $\sigma_\mathrm{on} = \tau_\mathrm{on}/3$, $\sigma_\mathrm{off} = \tau_\mathrm{off}/3$ for all pumps. In this configuration, in the volume $x,y \in [-4,4]$ cm, $z \leq 89$ cm , the mean flow remains small compared to the fluctuations, and the latter are satisfactorily isotropic as shown in Table \ref{table:hydro}. With a Taylor-scale Reynolds number of $Re_\lambda\approx 300$, the background turbulence is well-developed at the Taylor scale, with typical coherent eddy size of order $L = 6.6$ cm. Its intensity decreases exponentially with height over a characteristic length $L_d = 28.4$ cm. Turbulence will therefore decrease significantly as the interface travels up throughout each experiment.

\renewcommand{\arraystretch}{1.2}
\begin{table}[]
\normalsize
\begin{tabular}{ccc}
\hline\hline
Observable & Value from \citep{laplace_etude_2022} ($z=89$ cm) & Present study, at the interface  \\ \hline
$\sqrt{\langle {u'_{x}}^2\rangle}$                                        & 3.9 cm\persec  & similar or slightly lesser  \\ 
$\sqrt{\langle u_z^{\prime 2}\rangle}$                                    & 3.9 cm\persec  & lesser  \\ 
$|\langle u_x\rangle| / \sqrt{\langle u_x^{\prime 2}\rangle}$               & 0.03           & larger  \\ 
$|\langle u_z\rangle| / \sqrt{\langle u_z^{\prime 2}\rangle}$               & 0.09          & larger  \\ 
$\sqrt{\langle u_x^{\prime 2}\rangle} / \sqrt{\langle u_z^{\prime 2}\rangle} - 1$ & $0.02$  & much larger  \\ 
$Re_\lambda$                                                              & 300             & lesser  \\ 
$Re$ & 2,300 & lesser \\
$L$                                                                       & 6.6 cm          & -- \\ 
Energy dissipation rate $\epsilon$                                        & $5.0 \times 10^{-4}~\mathrm{m^2\!\cdot\!s^{-3}}$ & --
\\ \hline\hline
\end{tabular}
\caption{Characteristics of the flow induced by the RASJA in a homogeneous fluid (pure water).The first column reports measurements by \citet{laplace_etude_2022} at $z = 89$ cm. The second column reports qualitative estimates of the way the flow characteristics are affected by the slightly larger $z$, less powerful forcing, and the presence of a density interface. Brackets indicate horizontal and time averages and primed quantities refer to fluctuations with respect to the respective averages. $L$ is estimated from the integral of the velocity fluctuations autocorrelation.}
\label{table:hydro}
\end{table}

The base flow studied here differs somewhat from the work of \citet{laplace_etude_2022}: the pump power is set 25\% smaller (since strong forcing destroys the interface very quickly) and the region of interest is larger in the $xy$ plane ($x,y\in[-12, 12]$ cm) and farther up, ranging from to $z = 85$ to $z = 105$ cm as the interface is displaced upwards with mixing. These are the only changes in the \textit{control} parameters. In addition, the density interface affects significantly the flow, as discussed below. Nevertheless, the modification of the base flows it induces constitutes a time-dependent \textit{response} observable (the interface itself evolves with time) rather than a control parameter. In the present study, we focus solely on the properties of mixing itself, in relation with density fields, initial stratification and prescribed forcing only. This analysis does not require to precisely know the flow characteristics.

Still, we briefly discuss here the qualitative effects of stratification and of the modified forcing on the flow. Since the observation window is farther away from the forcing which itself is weaker, the rms of the velocity fluctuations and the Reynolds number are expected to be smaller. Isotropy of the base flow typically improves with height (and most likely with decreasing forcing power) \citep{laplace_etude_2022}. The density interface is however expected to have a contrary and dominant effect. In the case of stratified grid turbulence, enhanced mean circulations were reported \citep{poulain_etude_2020}, as well as an increase of horizontal velocity fluctuations coupled with a sharp reduction in the vertical velocity fluctuations \citep{poulain_etude_2020, Hannoun_Fernando_List_1988}. The interface was found to act qualitatively like a rigid flat plate and to affect strongly the largest scales of the flow \citep{Hannoun_Fernando_List_1988}. We summarise these considerations in Table \ref{table:hydro}.

\subsection{Parameter space}

\begin{figure}
    \centering
    \includegraphics[width=0.75\linewidth]{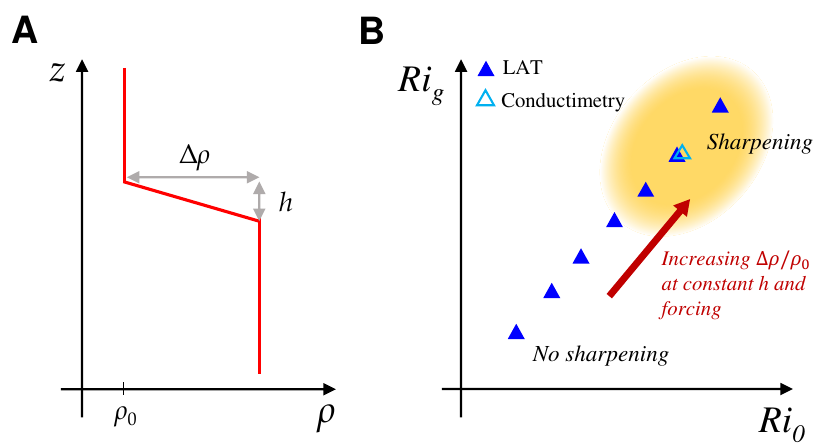}
    \caption{\textbf{(a)} Schematic of an ideal density interface. \textbf{(b)} Qualitative representation of the explored parameter space. The yellow shaded area indicated the region where interface sharpening was observed.}
    \label{fig:paramspace}
\end{figure}

An ideal density interface as depicted in Figure \ref{fig:paramspace}A is characterised by only two parameters, its thickness $h$ and its relative density jump $\Delta \rho/\rho_0$. Combined with the other experimental parameters, four control dimensionless numbers arise:
\begin{equation}
    Re_F = \frac{u^\prime L}{\nu},~~Sc = \frac{\nu}{D},~~Ri_0 = \frac{\Delta\rho gL}{\rho_0u^{\prime 2}},~~Ri_g = \frac{\Delta\rho gL^2}{\rho_0hu^{\prime 2}}.
\end{equation}
Here $L$ and $u^\prime$ refer to the integral length and the velocity rms of the base flow below (or without) the density interface. This definition ensures that the dimensionless numbers rely only on the initial and boundary conditions, rather than on the time evolution of the flow. \citet{Hannoun_Fernando_List_1988} made a similar choice, and showed that $u^\prime$ is unaffected by the interface at distances from the interface larger than $L$. $\nu$ and $D$ are the kinematic viscosity and molecular diffusivity of sodium chloride, respectively. Along with the Reynolds number of the forcing flow $Re_F\sim10^3$ and the Schmidt number $Sc = 7\times 10^2$, the bulk Richardson number $Ri_0$ compares gravitational potential energy to turbulent kinetic energy. It is independent of the specific initial shape of the interface and hence is appropriate to predict the steady state in a vertically infinite system. On the other hand, the gradient Richardson number $Ri_g$ compares only the initial density gradient $\Delta\rho/h$ to the characteristics of the forcing and as such is associated with the transient regime, when turbulent eddies have not yet managed to reach the upper layer and therefore are still unaffected by the total density jump $\Delta\rho$.

Here, we carried out a set of eight experiments with the upper layer density $\rho_0 = 1\,000 \pm 2$~kg\permcubed in all cases and the initial lower layer density $\rho_l = \rho_0 + \Delta\rho$. The RASJA is actuated with the same parameters for all experiments, starting at $t=0$ when the interface is at $z_i(t=0)\approx 92$ cm, and continuing during circa 25 minutes or until the interface has travelled out of the measurement volume. $Re$ and $Sc$ are therefore constant. While creating the interface we tried to keep $h$ as constant as possible from an experiment to another, in such a way that in the parameter space we explore is similar to a line on the ($Ri_g, Ri_0$) plane (Fig. \ref{fig:paramspace}B) parametrised by $\Delta\rho/\rho_0$. We estimate that $h= 2.5\pm 1$ cm for all experiments.

\subsection{Density measurements}

In seven experiments the fluid is monitored using two HD black and white cameras (MQ022MG-CM by Ximea) whose respective optical axes are perpendicular to the sides of the tank and to each other (see Fig. \ref{fig:setup}). Two LED panels are placed against the two faces opposite to the cameras. Cameras are synchronously triggered by a function generator. The salty-phase is acidified to pH $\sim 5.3$ to prevent calcium carbonate precipitation and ensure transparency. We carry out light-attenuation measurements (LAT), a technique that has been tried and tested for 30 years for geophysical or stratified flows \citep{holford_measurements_1996, allgayer_application_2012, dossmann_mixing_2017}. The dense layer is dyed with E133 before any turbulence is forced or dilution is made. Since the dye and salt diffusions are both much slower that momentum diffusion, they remain proportional to each other at the relevant flow scales. The diffusion coefficients of E133, Na$^+$ and Cl$^-$ are all similar ($\approx D$) and much smaller than $\nu$ ($Sc = 7\times10^2$). The density of the dense, salty phase is itself a linear function of the salt concentration. In LAT, the absorbance $A_{2/1} \equiv \log_{10}(I_2/I_1)$ of the solution is measured, where $I_1$ and $I_2$ are the intensity of the light passing through the water tank at a given position in two different flow states 1 and 2, measured by a given pixel of one camera. They are related to the density via Beer-Lambert law extended to polychromatic light:
\begin{equation}
    A_{2/1} = \log_{10}\left(\frac{\kappa e^{\tilde{\beta}(\rho_1-\rho_0)} + 1 - \kappa}{\kappa e^{\tilde{\beta}(\rho_2-\rho_0)} + 1 - \kappa}\right)
    \label{eq:linexp-abs}
\end{equation}
where $\tilde{\beta}$ and $\kappa$ are fitting parameters, $\rho_0$ is the density of fresh water and $\rho_1$ and $\rho_2$ are the densities of states 1 and 2 integrated along the optical path. The origin of equation \eqref{eq:linexp-abs} is explained in Appendix \ref{sec:beerlam}.

In one additional experiment with $\Delta\rho = 88$ kg\permcubed and the same forcing, instead of doing LAT we measured the conductivity $\sigma$ of the solution with a conductivity sensor (LFS1K0.1305.6W.B.010-6) from Innovative Sensor Technology which was automatically moved up and down periodically across the interface, from $z = 87$ cm to $z = 107$ cm with a period of 34 s. The probe was $12.9\times5.5\times0.65$ mm in size and was attached to a support only a centimetre wide, so as to minimally perturb the flow. In the sole presence of sodium chloride and at pH $~7$, there is a monotonic, bijective relation between the conductivity and the density (via the sodium chloride content), therefore these measurements provide an independent proxy of the density field.

\begin{figure}
    \centering
    \includegraphics[width=\linewidth]{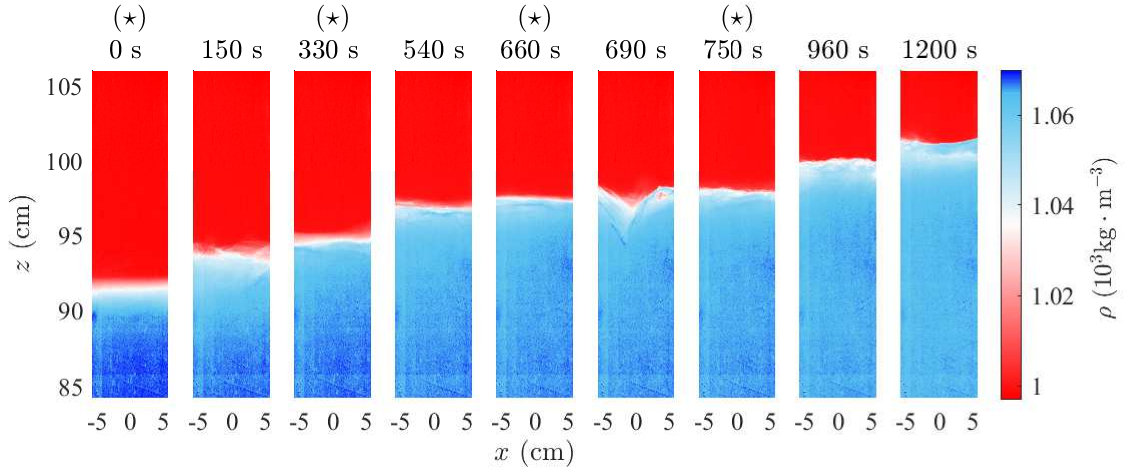}
    \caption{Density maps seen in LAT by one camera for $\Delta\rho = 73$ kg\permcubed. Strong mixing event alternate with calm phases (marked with $\star$) for which an interface position is defined. The upper layer is quiet, just nibbled. The lower layer is mostly homogeneous except close to the interface, and its average density decreases.}
    \label{fig:density_maps}
\end{figure}

\section{Interface displacement under mixing}\label{sec:displ}

\subsection{Stratification profiles}

\begin{figure}
    \centering
    \includegraphics[width=0.85\linewidth]{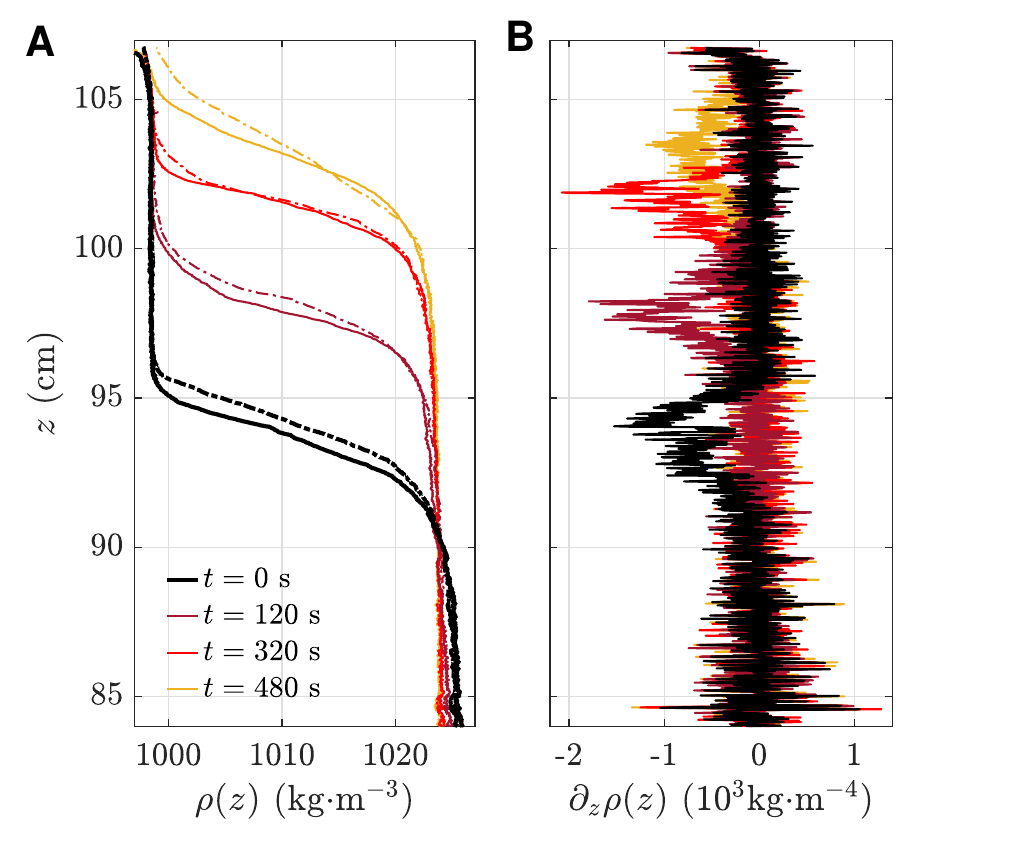}
    \includegraphics[width=0.85\linewidth]{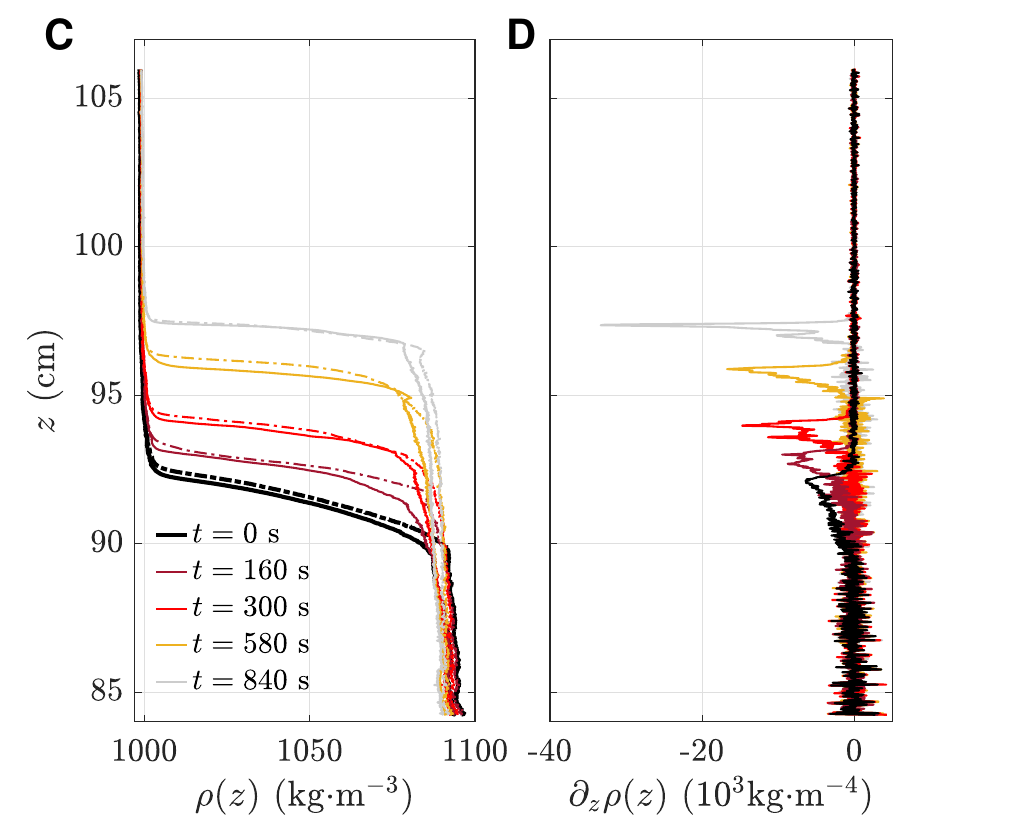}
    \caption{\textbf{(a, c)} Average density and \textbf{(b, d)} density gradient profiles for several times after the start of the forcing, indicated in plots. The top plots \textbf{(a, b)} correspond to $\Delta \rho=28$ kg\permcubed. The bottom plots \textbf{(c, d)} correspond to $\Delta\rho = 106$ kg\permcubed. Solid and dashed lines represent the two cameras.}
    \label{fig:profiles}
\end{figure}

\begin{figure}
    \centering
    \includegraphics[width=\linewidth]{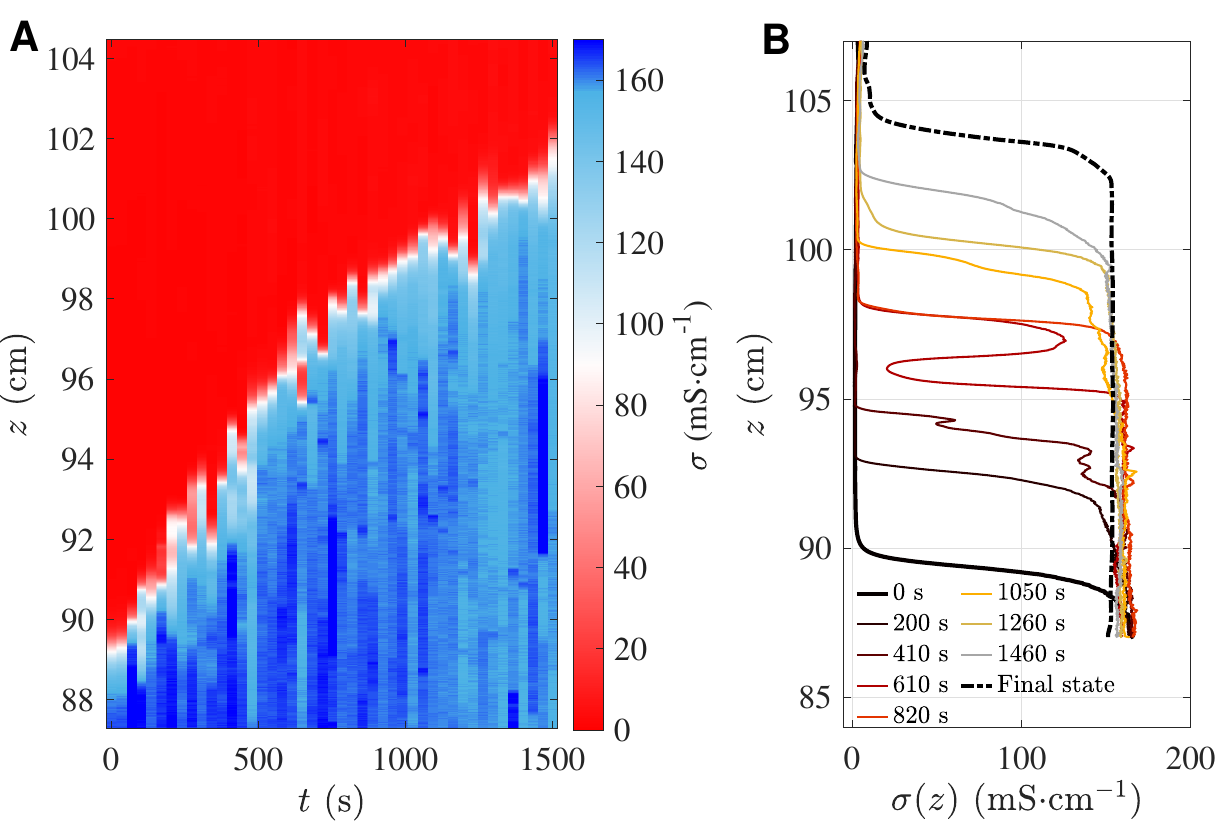}
    \caption{\textbf{(a)} Hovmöller diagram of conductivity and \textbf{(b)} vertical profiles of the conductivity at times indicated in plot, for $\Delta\rho = 88$ kg\permcubed. }
    \label{fig:hovmoller_cond}
\end{figure}

Typical density maps obtained with LAT are reported in Fig. \ref{fig:density_maps}. A general feature observed for all interfaces is that very strong mixing events (e.g. $t = 690$ s) alternate with much calmer phases (marked with $\star$). The density remains uniform and constant in the upper layer, in agreement with the well-known observation that the interface acts as a barrier for turbulence in such a way that the upper layer is undisturbed, but nibbled from the bottom \citep{rouse_diffusion_1955, gostiaux_diffusion_2013}. The lower phase is relatively homogeneous from a few centimetres below the interface, but its average density decreases as mixing progresses and buoyancy is pulled out of the upper layer and transported to depths (see the shades of blue in Fig. \ref{fig:density_maps}). When the interface looks quiet enough, it makes sense to average $\rho$ along $x$. Note that owing to parallax effects (see Fig. \ref{fig:setup}), the average we take is weighted: the light received by pixels on the sensor sides will be more absorbed due to longer optical paths than at the middle. Nonetheless, if the interface is quiet and uniform this shouldn't matter.

The horizontal averaging of density maps lead to the density profiles $\rho(z)$ and their respective gradients, shown in  Fig. \ref{fig:profiles} for two experiments ($\Delta \rho=28$ kg\permcubed and 106~kg\permcubed\!). Three features are observed in the evolution of these profiles. First, the progressive upward displacement of the interface which is caused by the mixing induced by turbulent eddies. Second, the displacement is associated with a slow yet measurable decrease of the density in the lower layer. Third, the density gradient exhibits two different behaviours in the two experiments. For $\Delta \rho=28$ kg\permcubed, a progressive thickening of the interface and decrease of the maximum gradient is measured, while the interface clearly sharpens for $\Delta \rho=106$ kg\permcubed. The observed sharpening mechanism at larger $\Delta \rho $ is discussed in the next section.

The density map obtained from the conductivity experiment at $\Delta \rho=88$ kg\permcubed  reported in Fig. \ref{fig:hovmoller_cond} shows similar features. Strong mixing events are observed (e.g. overturn at $t = 610$~s with very unstable stratification), alternating with calmer phases (e.g. $t = 920 - 1000$~s where the interface position seem to increase smoothly with time). The density is also uniform in the upper layer. However, the lower layer is less homogeneous in this case compared to Fig. \ref{fig:density_maps}, consistently with the low-pass filter effect of LAT which averages absorption along the optical paths. The average conductivity in the lower layer equally decreases.

\subsection{Mixing rate} \label{sec:displ-rate}

\begin{figure}
    \centering
    \includegraphics[width=\linewidth]{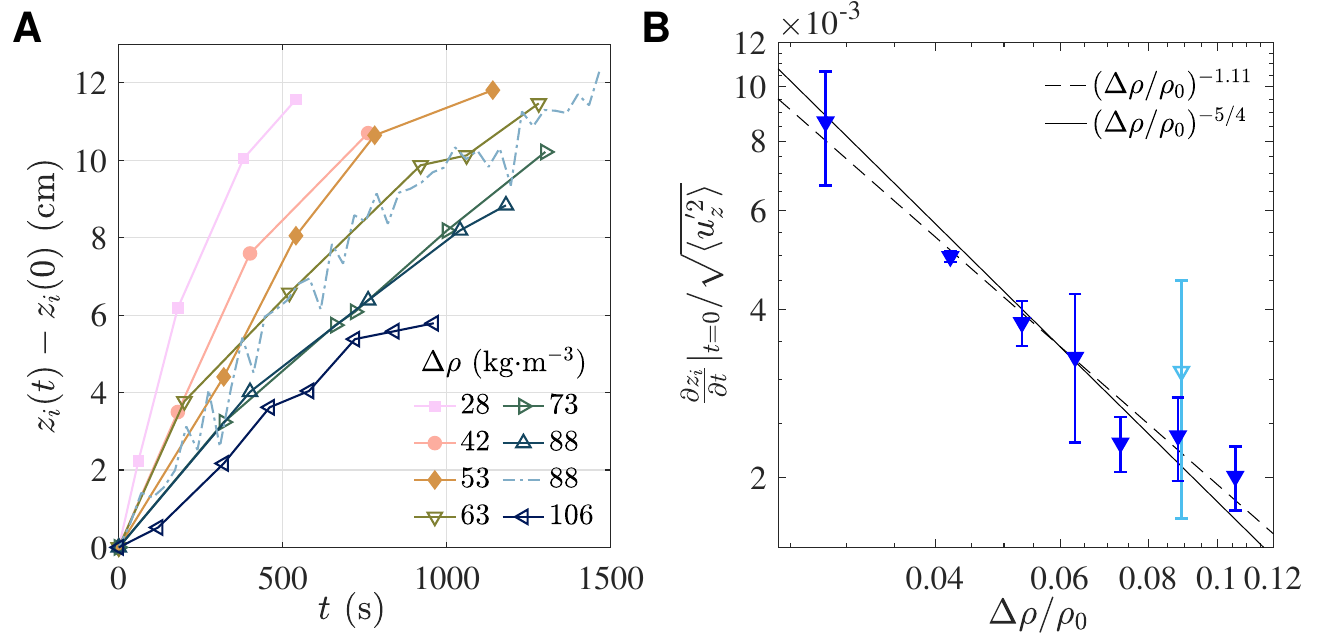}
    \caption{\textbf{(a)} Displacement of the interface over time for various density jumps $\Delta\rho$. The dash-dotted line for $\Delta\rho=88$ kg\permcubed corresponds to the point-like conductivity measurements. \textbf{(b)} Initial interface velocity as a function of the relative density jump $\Delta\rho/\rho_0$. It is consistent with a ($-5/4$)-power-law (solid line), the best fit exponent is $-1.11\pm0.5$ (dashed line). The lighter, empty symbol corresponds to the point-like conductivity measurement. Data are non-dimensionalised by the control parameter $\sqrt{\langle u_z^{\prime2}\rangle} = 3.9$ cm\persec from the non-stratified case.}
    \label{fig:displacement}
\end{figure}

A useful measure to quantify mixing and be able to parametrise it in global circulation models is the so-called \textit{mixing rate}. A first way to define this rate is to consider the velocity at which the interface is displaced up. The position of the quiet interface $z_i$ is determined from the density gradient profile: we first filter it and then calculates $z_i$ as an average of all $z$ weighted by $(\partial_z\rho)^3$. This method was chosen to limit sensitiveness to outliers (compared to a simple search for the maximum of $\partial_z\rho$), include the asymmetry of the density profiles, which often feature a tail towards large $z$ (not captured by a mid-density criterion), and to be insensitive to the measurement window size. $z_i$ is plotted against time for various $\Delta\rho$ in Fig. \ref{fig:displacement}a. LAT data seems to shows two different regimes: for $\Delta\rho \leq 50$ kg\permcubed (filled symbols), the velocity significantly decreases with time/$z$-coordinate, whereas is is relatively constant for $\Delta\rho \geq 60$ kg\permcubed in spite of the decreasing forcing intensity. It is not possible to tell from these measurements alone whether these are two qualitatively different regimes, or if the measurements in the second case are just to short to observe the slow down which could happen at yet higher interface displacement. In any case, the larger $\Delta\rho$ the slower the displacement. For comparison, we also provide the interface position inferred from the conductivity measurements (using a mid-density criterion). These measurements are instantaneous and point-like, therefore very sensitive to the specific waves experienced by the probe. They show an interface moving faster than in the LAT counterpart, but they nevertheless agree in trend with the LAT. With the interface having reached a higher altitude in this experiment, a slowdown is observed even with $\Delta\rho = 88$ kg\permcubed, supporting the second above-mentioned hypothesis.

To determine the initial interface velocity $\partial_tz_i(t=0)$ in spite of the sparse and noisy data, we fit the LAT curves either with 1st or with 2nd order polynomials excluding the last two points, and estimate the fit error using a bootstrap method with parametrised Gaussian noise. Fig. \ref{fig:displacement}b shows $\partial_tz_i(t=0)$ as a function of $\Delta\rho/\rho_0$, with error bars accounting for the choice of fitting polynomial order and the fit error. The same procedure is applied to the conductivity measurement (lighter empty symbol), with the sensitivity on the fitting window included in the error bars. \citet{rouse_diffusion_1955} found a relation that implies a scaling law $\partial_tz_i\propto \Delta\rho^{-5/4}$, although in a different setup (oscillating grid turbulence, interface maintained at a constant height). \citet{venaille_role_2014} too found a similar scaling law with an exponent in the estimated range $[-5/2, -1]$ (best-fit: $-5/4$), again in the case of grid turbulence but with a moving interface. The present results are compatible with this scaling ($R^2 = 0.98$), with a best fit exponent $-1.11\pm0.5$. They suggest that this scaling law could be valid in a more general framework or at least only little sensitive to the forcing type, making a step towards the generalisation of experimental results to more realistic systems.

\section{Interface sharpening}\label{sec:lay}

\begin{figure}
    \centering
    \includegraphics[width=0.28\linewidth]{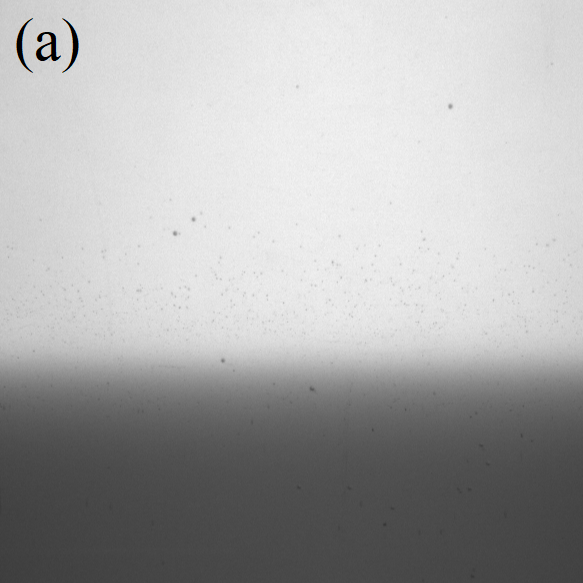}
    \hspace{0.02\linewidth}
    \includegraphics[width=0.28\linewidth]{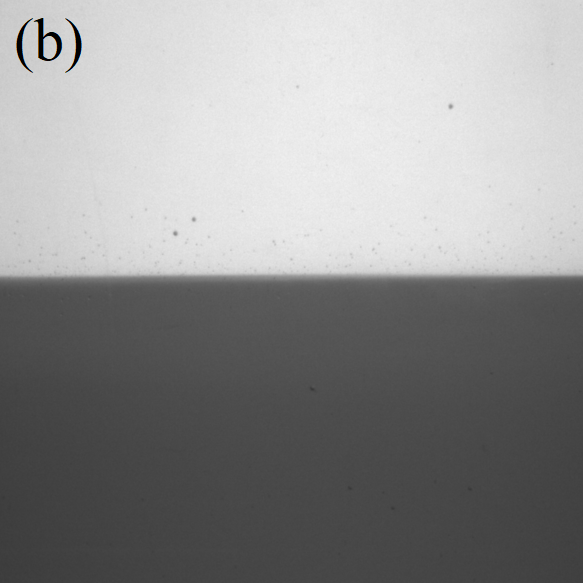}
    \caption{Raw picture of an interface with $\Delta\rho = 103~\mathrm{kg/m^3}$ \textbf{(a)} Initial state and \textbf{(b)} final state after 300~s of turbulent forcing and 60 s of rest. Interface sharpening is visible. }
    \label{fig:layering}
\end{figure}

\begin{figure}
    \centering
    \includegraphics[width=\linewidth]{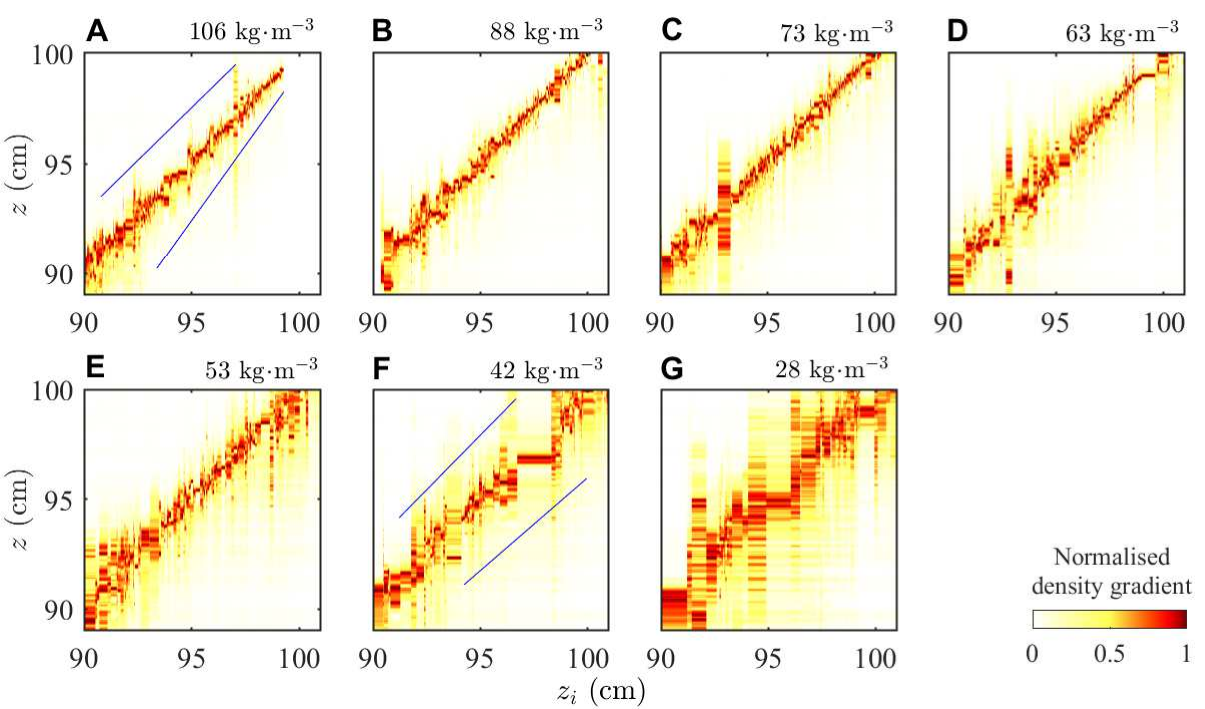}
    \caption{$N^2 = -g(\partial_z\rho)/\rho$, normalised by its maximum throughout the given experiment, against altitude $z$ and the position $z_i$ of the interface at this time, for the seven LAT experiments ($\Delta\rho$ indicated in plot).
    Blue lines are guides only. \textbf{(a)-(c)} exhibit a thinning of the interface characteristic of sharpening. \textbf{(f)-(g)} don't. \textbf{(d)-(e)} are intermediate.}
    \label{fig:layering_full}
\end{figure}

In the experiments with the largest values of $\Delta\rho$, the interface visually sharpens in the course of a mixing event, as shown in figure \ref{fig:layering}. This observation is indicative of the layering mechanism which has been characterized in previous experiments in stratified flows \citep{ruddick_formation_1989,park_turbulent_1994,holford_turbulent_1999} and reviewed by \citet{thorpe_layers_2016}. Through layering, a linear stratification is first destabilized by the appearance of small density steps, which then sharpen until the final density profile is stairs-like. In the present case, the initial profile already consists in a single sharp step, and we could observe the second phase of the layering mechanism where this step is made yet sharper. However, It is difficult to establish a single scalar metric for sharpening, because the interface thickness is not well defined when intense mixing events generate waves and lead to non-monotonic and/or asymmetric density profiles. Therefore, we study the full map of an intensive quantity instead: the square of Brunt-Väisälä frequency $N^2(z,t) = -g(\partial_z\rho(z,t))/\rho_0$. The larger $N$, the stronger the interface. $N^2$ is plotted in Fig. \ref{fig:layering_full}a-g against the vertical coordinate $z$ and the monotonically increasing interface position $z_i(t)$ (horizontal axis) defined in section \ref{sec:displ-rate}.

\begin{figure}
    \centering
    \includegraphics[width=0.85\textwidth]{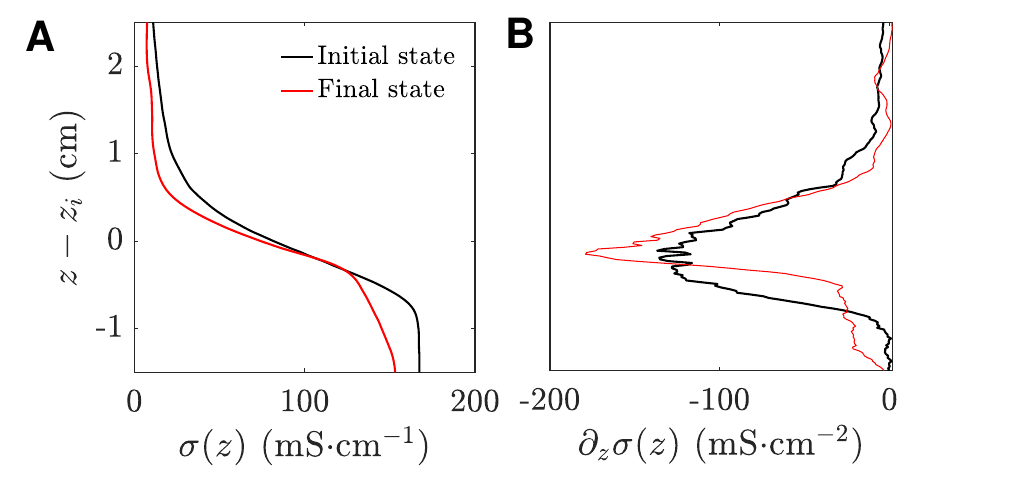}
    \caption{Interface sharpening observed at $\Delta\rho = 88$ kg\permcubed by conductivity measurements. \textbf{(a)} Conductivity and \textbf{(b)} conductivity gradient measured before and after 1500 s of turbulent forcing and $\sim$ 60 s of rest.}
    \label{fig:layering_cond}
\end{figure}

For large $Ri_0$ ($\Delta\rho \geq 70$ kg\permcubed\!), the large-gradient region (orange and red shades) narrows significantly as the interface travels up, in proportions that are significant regardless of parallax or refractive index gradient effects as the guiding lines of Fig. \ref{fig:layering_full}a show. Notably, parallax tends to widen the apparent large-gradient region when it is away from the optical axis ($z = 95$ cm), whereas graphs \ref{fig:layering_full}a-c show a constant thinning of that region throughout the experiment. On the other hand, no such sharpening is observed at $\Delta\rho \leq 50$ kg\permcubed but the large-gradient area thickens instead, like one expects in low $Ri_0$ regime. From the data presented here, density jumps $\Delta\rho \in [50, 70]$ kg\permcubed seem to belong to some intermediate, transitional state with neither thinning nor thickening of the interface. It is possible that they exhibit one of the two aforementioned behaviours but on time scales longer than that of our experiment.

We confirm these observations by conductivity measurements at $\Delta\rho = 88$kg\permcubed, where LAT showed that sharpening is to be expected. These measurements are completely unaffected by any parallax or refractive index effect. The conductivity profile and vertical conductivity gradient profiles are compared before mixing and after mixing and $\sim 60$ s of rest in Figure \ref{fig:layering_cond}. After mixing, the maximal conductivity gradient is enhanced by 32 \% while the peak's full width at half-maximum is reduced by 35 \%.

The present findings can be examined in the light of the relation between the dynamical structures and the stratification budget presented in the direct numerical simulations of a turbulent Couette flow in \citep{zhou_diapycnal_2017}. The evolution of the stratification is analysed in three regimes controlled by $Sc$ and $Ri_0$. At low $Ri_0$, the destabilising mechanism are intermittent shear-induced local overturns that mixes the interface and decrease its sharpness (turbulent \textbf{T} regime). At larger $Ri_0$ and low $Sc$, stratification suppresses turbulent motions at the interface and decaying interfacial waves are observed (laminarising \textbf{L} regime). Increasing $Sc$ from the \textbf{L} regime leads to the generation of Holmboe wave structures, scouring either sides of the interface. In this regime, the isopycnal surfaces $A_s$ are enlarged by the scouring effect above and below the mean interface position, while isopycnals in the median region are hardly perturbed. The effective salt diffusivity $\kappa_e (z)$ being proportional to $A_s^2$, the above and below regions are well mixed while the mean interface is protected, enhancing the interface sharpness. We stress that no direct evidence of such dynamical structures (e.g. resolved interfacial waves or buoyancy flux measurements) is available in the present experiments. Therefore, this comparison should be understood as a qualitative interpretative framework rather than a demonstration of the underlying mechanisms. Additionally, in the present experiments the forcing is not induced by a steady shear flow, but by non-stationary intense mixing events induced by self-interacting turbulent jets impinging on the pycnocline. Despite the fluctuating nature of the forcing flow where strong mixing events at the interface alternates with quieter, laminar displacements, the long term behaviour of the interface is similar to the observations of \citep{zhou_diapycnal_2017}, resembling the turbulent state at low $Ri_0$ and the Holmboe state at high $Ri_0$ and indicating that the layering vs shear-induced behaviour is robust to the forcing mechanism. Note that the laminarising regime observed at low Schmidt number ($Sc=7$) is not observed in the present experiments where $Sc = 7\times10^2$.

\section{Potential Energy budget}\label{sec:energy}

\begin{figure}
    \centering
    \includegraphics[width=0.85\linewidth]{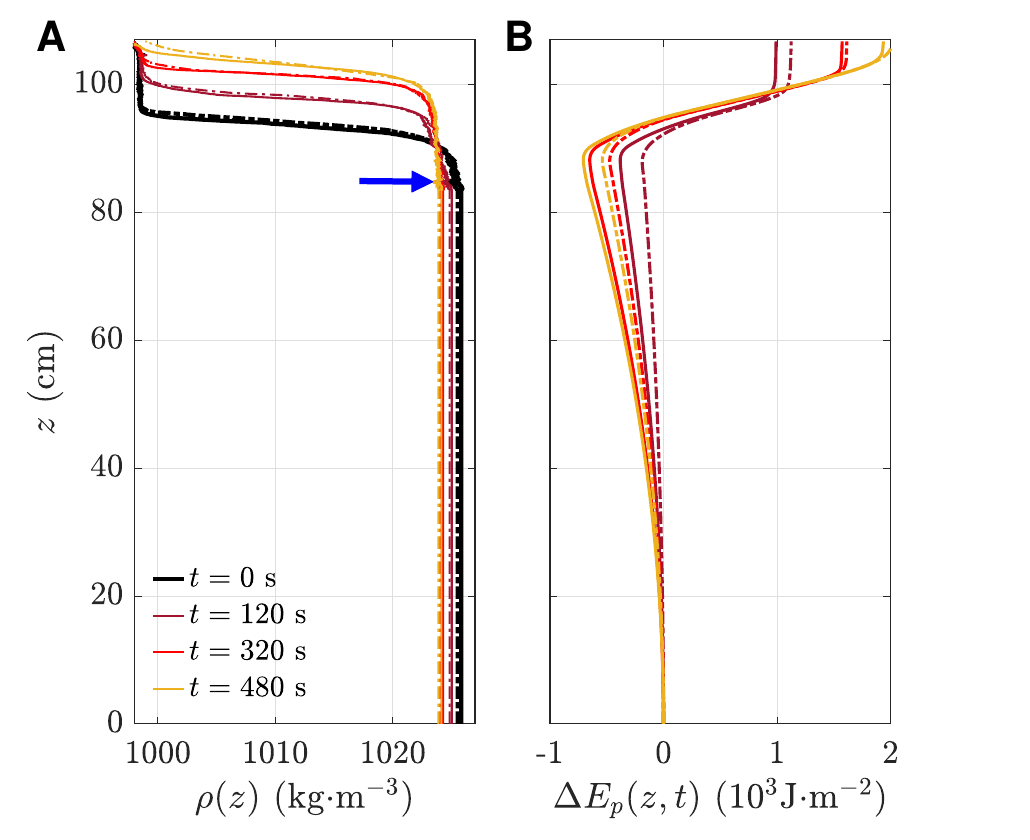}
    \includegraphics[width=0.85\linewidth]{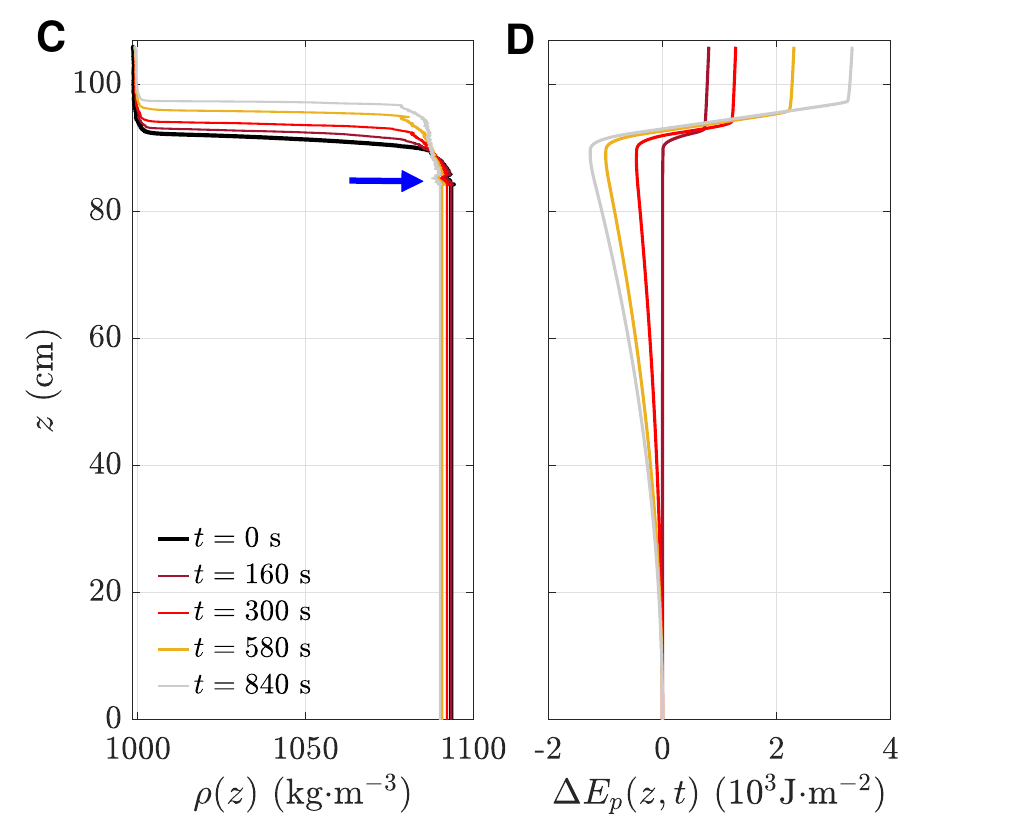}
    \caption{\textbf{(a, c)} Extrapolated density profiles and \textbf{(b, d)} corresponding cumulative potential energy $E_p(z,t)$ at selected times indicated in the plots. The top plots \textbf{(a, b)} correspond to $\Delta \rho=28$ kg\permcubed. The bottom plots \textbf{(c, d)} correspond to $\Delta\rho=106$ kg\permcubed. Solid and dashed lines represent the two cameras. For clarity the second camera is omitted in \textbf{(c, d)}. The blue arrows demarcate the measurement window from the extrapolated zone.}
    \label{fig:profiles_energy}
\end{figure}

The density profiles taken at rest permit to quantify the evolution of the cumulative gravitational potential energy until $z$, $E_p(z,t)=g\int_0^z d^3\vec{r'} \rho(z',t)z'$. Note that at the instants when the interface is quiet for which we calculate $E_p$, there is no evidence of overturns and the density field appears (almost) statically stable. Therefore, $E_p(t, H)$ almost coincides with the background potential energy $E_{b}(t) = g\int_0^{H} d^3\vec{r'} \rho(z',t)\tilde{z}(\vec{r}')$, defined as the potential energy of the fluid after the adiabatic, incompressible reordering $\tilde{z}(\vec{r}')$ of the fluid parcels which establishes a statically stable density profile. Changes in $E_b$, by definition, correspond to \textit{irreversible} changes of potential energy caused by the (irreversible) mixing process. \cite{winters_available_1995, dossmann_experiments_2016, dossmann_mixing_2017}.

The variation of $E_p$ in the measurement field (with the origin taken at the bottom of the field) shows an increase for both thickening and sharpening scenarios. In order to evaluate the integrated $E_p(z,t)$ in the entire domain, density profiles from LAT are extrapolated to the base of the tank, assuming a constant density equal to the value at the bottom of the measurement window. This procedure is shown in Figures \ref{fig:profiles_energy}a and \ref{fig:profiles_energy}c for $\Delta \rho=28$ kg\permcubed and $\Delta \rho=106$ kg\permcubed, respectively. The extrapolation is supported by the intense turbulent mixing forced by the pumps that homogenise the bottom layer, up to the lower part of the measurement window (Figs. \ref{fig:density_maps}, \ref{fig:profiles}a and \ref{fig:profiles}c). We further quantify the error due to this approximation by estimating the total fluid mass from the extrapolated profiles and find variations of less than $0.1 \%$ over time. At each time, the change in cumulative potential energy $\Delta E_p(z,t) = E_p(z,t) - E_p(z,0)$ shown in Figures \ref{fig:profiles_energy}b and \ref{fig:profiles_energy}d exhibits a decreasing vertical trend in the lower layer owing to the density decrease by mixing with lighter fluid parcels. The rise of the interface overcompensates this decrease and leads to a strong positive variation of $\Delta E_p(z,t)$ in the narrow region surrounding the interface. This behaviour confirms that mixing processes participate to the rise of the centre of mass of the fluid. We show that this is the case for both scouring and sharpening scenario by examining the evolution of the potential (almost background) potential energy of the entire fluid column, $\Delta E_p(H,t)$.
\begin{figure}
    \centering
    \includegraphics[width=0.55\linewidth]{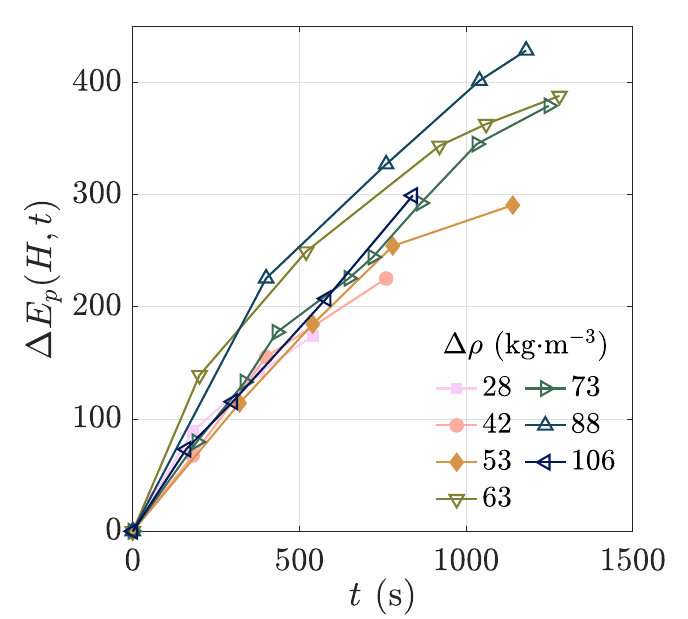}
    \caption{Change in the total potential energy (approximately equal to the background potential energy), for several density jumps $\Delta\rho$ indicated in plot.}
    \label{fig:bpe}
\end{figure}
This quantity is displayed against time in Fig.~\ref{fig:bpe}, for the times where the fluid is at transient rest. All experiments display similar trends, with a fast $\Delta E_p(H,t)$ growth rate when the interface is low at early times and a deceleration of the growth rate at later times. Note that while the initial density difference varies by a factor of circa $4$ across experiments, this variation hardly affects $\Delta E_p$ changes in the course of the experiment.

The cumulative potential energy input results from the fraction of kinetic energy injected at the bottom of the stratified column that eventually contributes to mixing the fluid. In order to provide a quantitative evaluation of the mixing efficiency, that is the ratio of (background) potential to injected kinetic energy, a precise characterization of the turbulent energy dissipation profile in the presence of a background stratification is required. Future experiments relying on Lagrangian characterisation of the velocity field will allow such diagnostics.

\section{Concluding remarks} \label{sec:conc}
The present work addressed the question of the evolution of a two-layer stratification submitted to an intermittent, turbulent forcing induced by randomly activated jets. Throughout LAT and conductivity experiments, the gradient and bulk Richardson numbers were varied proportionally, revealing two regimes. For small density jumps, turbulent structures overturn the local stratification, enhancing the local turbulent diffusivity and weakening the stratification. The second regime at larger $Ri_0$ and $Ri_g$ corresponds to the layering process commonly observed in grid turbulence experiments and turbulent Couette flow simulations. Here, the sharpening mechanism is not sustained by a steady turbulent forcing, but by time-limited extreme turbulent events. The robustness of the sharpening process to the intermittency of the forcing mechanism can help to understand the numerous occurrences of step-like structures in lakes and oceans.

This paper relied only on the characterisation of the density fields. The characterisation of the flow by particle tracking velocimetry (PTV) is the object of current research, and another group is designing simulations of a similar system. These developments should help understand how intermittent flows are affected by stratification. Flow measurements would also be required to quantify the mixing efficiency, a key parameter in ocean general circulation models.

Last, although this work clearly remains on the side of idealised experiments, it makes one small step towards the bridging of the relevance gap \cite{caulfield_layering_2021} using a more realistic, isotropic remote type of forcing. In the future, more and more complexity could be incorporated using inputs from in situ measurements.

\bmhead{Acknowledgements}
The authors are grateful to Antoine Venaille (ENS-Lyon) for discussions.

\section*{Declarations}

\begin{itemize}
\item Competing interest: The authors declare no competing interests.
\item Ethics approval and consent to participate: The published results are purely technical and did not involve any human, animal or biological samples. All authors of this study have fulfilled the criteria for authorship, and responsibilities were agreed among the collaborators.
\item Data availability: The datasets reported in this publication are available from the corresponding authors upon  request.
\item Code availability: The post-processing codes used for this publication are available from the corresponding authors upon  request.
\item Authors contribution: YD and MB designed the research. NC carried out the LAT measurements. NC and YD analysed the data. HP carried out the conductivity measurements. NC and YD wrote the manuscript. All authors discussed the results and reviewed the manuscript.
\end{itemize}

\begin{appendices}

\section{Model for the extended Beer-Lambert law}\label{sec:beerlam}

\begin{figure}
    \centering
    \includegraphics[width=0.5\linewidth]{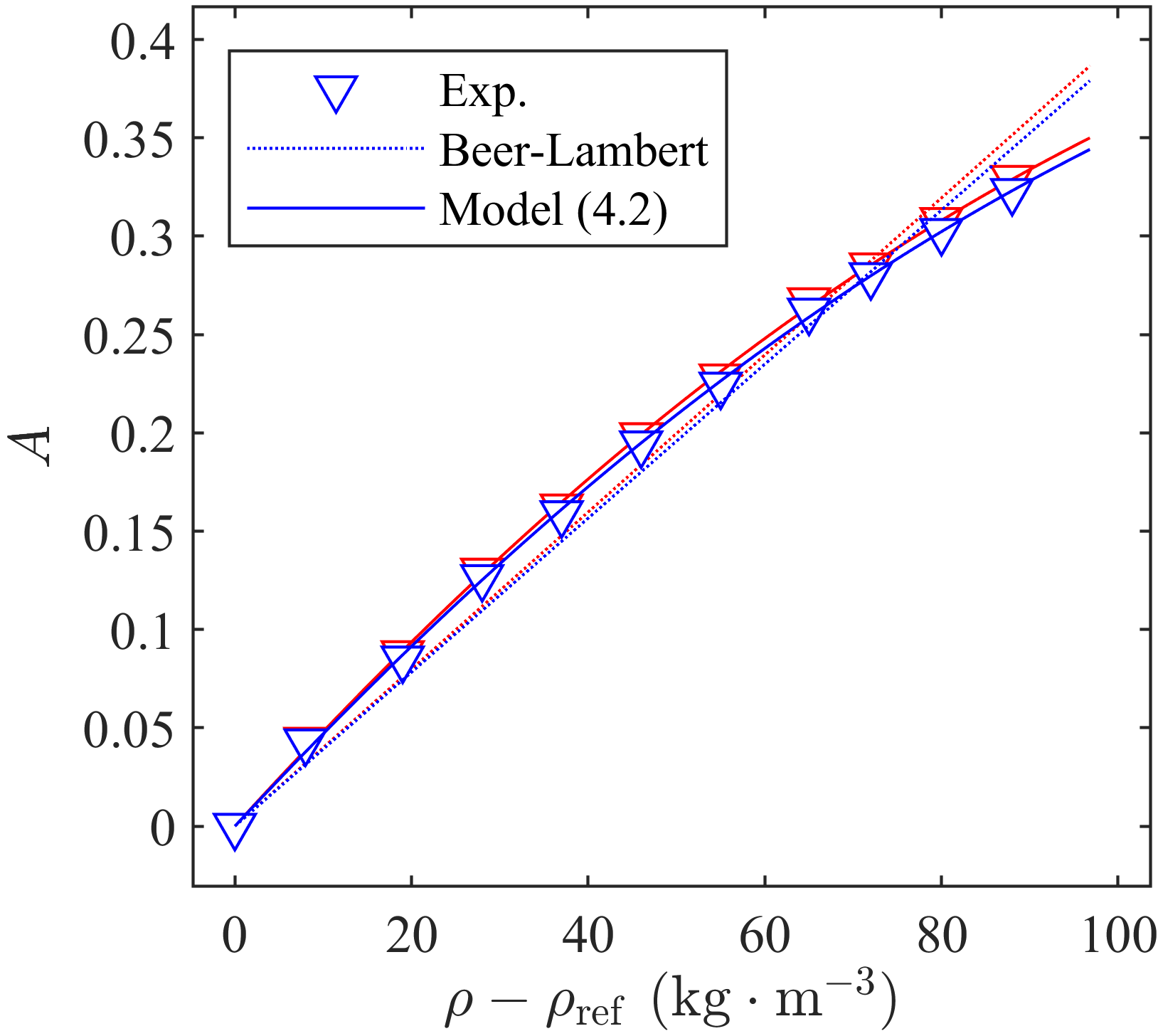}
    \caption{Absorbance to density relation. \textbf{Dashed lines:} linear fit (Beer-Lambert's law). \textbf{Solid lines:} linear-exponential model (\ref{eq:linexp-abs}) with $\kappa = 0.73$. Each colour corresponds to a camera.}
    \label{fig:calibration-absorbance}
\end{figure}

In the large-density-gradients, large-salt-concentration regime we focus on here, the equation of state of saline water is linear. Its density $\rho$ reads $\rho = \rho_0 + \alpha c_\mathrm{NaCl}$ which we can rewrite $\rho = \rho_0 + \tilde{\alpha}c_\mathrm{dye}$ where $\alpha$ is a known constant and $c_\mathrm{dye}$ and $c_\mathrm{NaCl}$ the dye and salt molar concentrations, respectively. We do not control accurately the salt-to-dye concentration ratio, therefore $\tilde{\alpha}$ is not known quantitatively. However, experiments with various $Ri_0$ are carried out in a row by successive dilutions, which ensures that $\tilde{\alpha}$ is a constant throughout the measurement campaign.

The usual absorbance-concentration relation is Beer-Lambert law, which states that the logarithm of the light intensity $I_2$ of each pixel normalised by the intensity $I_1$ of (any) other picture with varies linearly with $c$. Thus, $A_{2/1} \equiv - \log_{10}(I_2/I_1) = \tilde{\beta}(\rho_2 - \rho_1)$ where $\tilde{\beta}$ is a constant throughout the measurement campaign. However, the LED panels used are polychromatic and Beer-Lambert law must be modified accordingly. Let us assume that the lighting intensity $\mathcal{I}_0(\lambda)$ is constant in the wavelength range $[\Lambda_1, \Lambda_2]$ and is zero elsewhere, and the molar attenuation coefficient of the dye $\varepsilon(\lambda)$ to have a similar behaviour but in the range $[\lambda_1, \lambda_2]$. Then,
\begin{equation}
    I_{1,2} = \int \mathrm{d}\lambda \,\mathcal{I}_0(\lambda) \,10^{-\varepsilon(\lambda)l c_{1,2}}
\end{equation}
yields
\begin{equation}
    A_{2/1} = \log_{10}\left(\frac{\kappa e^{\tilde{\beta}(\rho_1-\rho_0)} + 1 - \kappa}{\kappa e^{\tilde{\beta}(\rho_2-\rho_0)} + 1 - \kappa}\right)
    \label{eq:linexp-abs-appendix}
\end{equation}
with $\tilde{\beta}$ as above, $l$ a length, $\kappa = (\lambda_2 - \lambda_1)/(\Lambda_2 - \Lambda_1)$ and $\rho_0$ the density of fresh water (\citet{cenedese_concentration_1998} used this relation too in a less general case).

Throughout the series of successive experiments, calibration pictures of the tank filled with a homogeneous, saline layer are taken, and the density of the fluid is measured with a densimeter. Then the fitting parameters $\kappa$ and $\tilde{\beta}$ are inferred. Fig. \ref{fig:calibration-absorbance} diplays the results of this procedure. The thick lines show that the model fits very well the entire data set with adjusted values $\tilde{\beta} = 7.08~\mathrm{g^{-1}\cdot cm^3}$ and $\kappa = 0.73$ falling within $15\%$ of our prior estimates based on order of magnitudes $\tilde\beta \approx 6$ and $\kappa \approx 0.72$. For reference, dashed lines show that Beer-Lambert law is not sufficient. 

However, more thorough analysis of the data reveals that the very good agreement \textit{on average} to the model above hides significant spatial inhomogeneity. In particular, due to the geometry of the measurement area (see Fig. \ref{fig:setup}), light rays coming from the sides, top and bottom cross the fluid on longer distances which influences $l$ and hence $\tilde{\beta}$. We therefore divide the pictures in 144 sections (each 100 pixels wide, i.e. $\approx$1.8 cm) and adjust the model parameters for each of them. This process corrects for the majority of the heterogeneity observed.

\end{appendices}

\bibliography{bibliography.bib}

\end{document}